\documentclass[12pt]{article}
\usepackage{mathtools}
\usepackage{empheq}
\usepackage{subcaption}
\usepackage[round]{natbib}
\usepackage[top=1.25in, bottom=1.25in, left=1.25in , right=1.25in]{geometry}

\def\btr{\begin{tabular}}       \def\etr{\end{tabular}}

\begin{document}

\def\bc{\begin{center}}         \def\ec{\end{center}}
\def\btr{\begin{tabular}}       \def\etr{\end{tabular}}
\def\btb{\begin{tabbing}}       \def\etb{\end{tabbing}}
\def\bi{\begin{itemize}}        \def\ei{\end{itemize}}  \def\i{\item}
\def\bd{\begin{description}}    \def\ed{\end{description}}
\def\bs{\bigskip}\def\ms{\medskip}\def\ss{\smallskip}
\def\sls{\bigskip\hrule\bigskip}
\def\slsc{\vspace*{-.75cm}\bc\underbar{\hspace*{5.5in}}\ec\vspace*{-.75cm}}
\newcommand{\Frame}[1]{{\footheight=0.0in\fboxrule=1.5pt\fboxsep=1ex\fbox{\btr{c}#1\etr}}}
\def\ns{\vspace*{-1.ex}}        \def\nss{\vspace*{-2.ex}}

\newcommand\ic{\mathrm{i}}
\newcommand\Real{\mbox{Re}} 
\newcommand\Imag{\mbox{Im}} 
\newcommand\St{\mbox{\textit{St}}} 
\newcommand\Rey{\mbox{\textit{Re}}}  

\newcommand{\dd}{\; \mathrm{d}}  

\def\ni{\noindent}


\title{\large\sc Efficiency of Fish Propulsion}

\author{{\normalsize A.P.~Maertens, M.S.~Triantafyllou \& D.K.P.~Yue}
\vspace{1cm}\\
{\small Center for Ocean Engineering, Department of Mechanical
Engineering}\\
{\small Massachusetts Institute of Technology}\\
{\small Cambridge, Massachusetts 02139, USA}
\vspace{0.5cm}\\
{\small {\em email:} maertens@mit.edu, mistetri@mit.edu, yue@mit.edu}}

\date{{\small \today}}

\maketitle

\begin{abstract}
It is shown that the system efficiency of a self-propelled flexible
body is ill-defined unless one considers the concept of {\em
quasi-propulsive efficiency}, defined as the ratio of the power
needed to tow a body in rigid-straight condition over the power it
needs for self-propulsion, both measured for the same speed. Through
examples we show that the quasi-propulsive efficiency is the only
rational non-dimensional metric of the propulsive fitness of fish
and fish-like mechanisms. Using two-dimensional viscous simulations and the concept of quasi-propulsive
efficiency, we discuss the efficiency two-dimensional undulating
foils. We show that low efficiencies, due to adverse body-propulsor hydrodynamic interactions, cannot be accounted for by the increase in friction drag.
\end{abstract}


\section{Introduction \label{sec:intro}}

Efficiency is defined as the ratio of useful work over expended
energy, measured over a specific time interval.  The useful work, in
a body moving at constant speed within a viscous medium, is the work
needed to overcome the resisting fluid forces (drag forces). This
work, however, cannot be measured except in very few, limiting
cases, because the drag coefficient of a self-propelled body depends
not only on its shape and speed, but also on the type of propulsor
used, and, in particular, the body-propulsor hydrodynamic
interaction. This is especially true for flexible bodies, where
propulsive forces are generated by body deformations that influence
significantly the drag force.

The propulsive efficiency is easy to define in the case of a
propulsor in isolation. For a propeller tested in a recirculating
tank, for example, a motor is used to drive the propeller, so that
the product of motor torque $Q$ times the rotational velocity
$\omega$ provides the expended energy per unit time (or 
power input $P_{in}$); the measured net thrust $T$ times the velocity $U$ of the
flow in the tunnel provides the useful work per unit time (or useful
power output $P_{out}$). Hence the efficiency $\eta$ is defined as:
\begin{equation}
\label{eq:eta}
\eta = \frac{P_{out}}{P_{in}}=\frac{T U}{Q \omega} \qquad \text{for an isolated propeller}. 
\end{equation}

In a self-propelled body moving at constant speed $U$, the total average
hydrodynamic force on the body must be zero, so using the definition of Eq. (\ref{eq:eta}), the net efficiency is $\eta_n=0$ if the net thrust of the entire body $T_{n}$ is used:
\begin{equation}
\label{eq:eta_n}
\eta_n = \frac{T_n U}{Q \omega}=0 \qquad \text{for a self-propelled body is steady state}.
\end{equation}
Under special
circumstances, one could still define a {\em propulsor
efficiency}, $\eta_p$, by separating the propulsor thrust $T_p$ from the body drag (one balancing the other in this case). For example, in a
ship propelled by a propeller equipped with force and torque
sensors, one could measure directly the force $T_p$ developed by the
propeller, as well as the torque $Q$ needed; then the equation
\begin{equation}
\label{eq:eta_p}
\eta_p = \frac{T_p U}{Q \omega}
\end{equation}
could be used directly to evaluate the propulsor
efficiency.  Additional losses, in the gears, transmission, etc.,
could be accounted for without much difficulty.  

For flexible self-propelled bodies, such as undulating
fish, where the distinction between thrust and drag cannot be made,
obtaining $T_p$ is much more challenging and can be seen as arbitrary.
It is still possible, in some cases, to estimate the thrust produced by
a swimming fish. Indeed, when the Reynolds number is sufficiently high 
and uncontrolled flow
separation effects are of limited extent, inviscid methods can be
used to provide an estimate of the power needed for propulsion, as
well as the developing thrust that must equal the resistance. This
can be quite accurate if separation effects, other than vorticity
shed from body edges and from fin trailing edges, are small, and
interaction of the body with shed vorticity is insignificant. For instance,
\citet{lighthill_large-amplitude_1971}, \citet{wu_hydromechanics_1971},
 \citet{drucker_locomotor_1999}, \citet{pedley_large-amplitude_1999}, 
\citet{wolfgang_near-body_1999}, and \citet{zhu_three-dimensional_2002}
employ inviscid methodologies to estimate the thrust generated 
and power expended by swimming fish. 

However, the main problem with this definition of efficiency is that, 
even in rigid bodies, such as ships and autonomous vehicles,
one is not interested in the propulsor efficiency, but the power
needed to sustain a certain speed. The efficiencies defined by equations
(\ref{eq:eta_n}) and (\ref{eq:eta_p}) are not appropriate metrics, unlike
the quasi-propulsive efficiency presented in section \ref{sec:eta_QP}. In
section \ref{sec:efficiencies}, we formally define three measures of
efficiency in the context of self-propelled bodies: the net propulsive 
efficiency, the propulsor efficiency and the quasi-propulsor efficiency.
In sections \ref{sec:naval} and \ref{sec:fish}, we show through 
examples that the quasi-propulsive efficiency
is the only rational measure of propulsive efficiency for a self-propelled
body in steady motion. Finally, we use this measure to quantify the
efficiency of two commonly used fish swimming gaits.


\section{Quasi-Propulsive Efficiency as a Performance Index \label{sec:eta_QP}}

The goal of propulsion optimization is set as follows: {\em For a
given shape and size vehicle, find the propulsor that will require
the least amount of power to drive the vehicle at a given speed
$U$.}

The reason why the problem is not set in terms of a {\em propulsive
efficiency} is that the propulsor interferes hydrodynamically with
the hull of the vehicle.  Hence, it is possible that a very
efficient propulsor (when tested detached from the vehicle) may
cause a large increase in the total drag when attached to the vessel
due to adverse hydrodynamic interference, and hence an increase in
the required thrust $T_p$. Then, although the propulsor efficiency is
high, the system efficiency is low because the torque, and hence the
fuel needed, may be excessive over another propulsor that may be
less efficient in isolation but does not affect the resistance.  In
summary, {\bf we intend to minimize the ``fuel'' consumption under
certain size and velocity constraints} and not the hydrodynamic
 efficiency of the system.

Still, it is better to deal with non-dimensional quantities, so as
to be able to derive scaling laws, and apply the fundamental
principles derived from one type of vehicle to another one. A possible
way of normalizing the power required to drive the motion of a fish or bird
 has recently been proposed
by \citet{bale_energy_2014} using the frequency, wavelength and amplitude of the motion. While the authors have shown that this non-dimensional power coefficient makes it possible to compare the efficiency of a wide range of swimming and flying animals, it is clearly not appropriate to compare the efficiency of different propulsion parameters, as the normalization depends on these parameters.

The power needed for propulsion is the result of complex
mechanical/hydrodynamic interactions and is difficult to
non-dimensionalize in a convenient way for scaling.  Hence, the
concept of {\em quasi-propulsive efficiency}, $\eta_{QP}$, is
employed, to reflect the original intent to minimize expended power:
{\bf For a vehicle with given towed resistance $R$ at a speed $U$,
find the propulsor that maximizes $\eta_{QP}$}, defined as:
\begin{equation}
\label{eq:eta_QP}
\eta_{QP} = \frac{R U}{P_{in}},
\end{equation}
where $P_{in}$ is the power required by the propulsor to drive the
vehicle at speed $U$ under steady-state conditions (zero total
hydrodynamic force). {\bf In the case of a flexible body, the towed
resistance must be measured or estimated in a straight
configuration, i.e. not allowing any bending of the body.} 
For example, in the case of a propeller, $P_{in}=Q \omega$ as
before, but $Q$ and $\omega$ must be measured under self-propulsion
conditions.

There are fundamental differences between the propulsive efficiency
of equation (\ref{eq:eta_p}) and the quasi-propulsive efficiency of
equation (\ref{eq:eta_QP}): First, in the ``useful'' power of equation
(\ref{eq:eta_QP}) one uses {\bf the towed resistance} of the vehicle
measured under steady towing conditions at speed $U$ and {\bf without a
propulsor} attached; hence, this definition does not suffer from any
ambiguity as to what the force should be.  Second, in (\ref{eq:eta_p})
all quantities used refer to the same (self propulsion) test; in
(\ref{eq:eta_QP}) the numerator refers to a towing experiment, while the
denominator to a self-propulsion experiment, conducted at the same
speed.

It is not difficult to see that, if we maximize the efficiency
$\eta_{QP}$, we simply minimize the expended power (since the
numerator is a constant), in agreement with the original intent. The
advantage of $\eta_{QP}$ is that the towed resistance captures the
essential hydrodynamic features of the specific hull, and can be
used to compare the performance of dissimilar vehicle shapes, and
for devising scaling laws.  An apparent disadvantage is that the
quasi-propulsive efficiency is not strictly an efficiency; hence it
is not necessarily less than one.  If the propulsor causes the
resistance of the ship to drop substantially -- for example by
reducing flow separation -- then the required power will possibly be
less than the power needed to tow the bare hull, resulting in a
value of $\eta_{QP}$ higher than $100\%$.


\section{Efficiency Measures in Self-Propulsion \label{sec:efficiencies}}

Let us consider the general case of a self-propelled body of mass $m$ moving with acceleration $a$ and velocity $U$ (both averaged over a period for periodic propulsion) along the $x$-direction. Considering the system \{body + propeller\} as a whole, the efficiency (referred to as net efficiency $\eta_n$) in its strictest definition is the ratio of the power output $P_{out}$ to the power input $P_{in}$: 
\begin{equation}
\label{eq:eta2}
\eta_n = \frac{P_{out}}{P_{in}}.
\end{equation}
The power output is given by the rate of change of kinetic energy (averaged over a period) of the body:
\begin{equation}
\label{eq:Pout}
P_{out}=\frac{\mathrm{d}}{\mathrm{d}t}\left( \frac{1}{2}m\,U^2  \right) 
=maU=T_nU,
\end{equation}
with $T_n$ the net thrust produced by the \{body + propeller\} system. This definition of efficiency is the same as that used for an isolated propeller.

Going back to the intuitive definition of efficiency, which is the ratio of useful work to total work, different configurations can be compared. A propeller in isolation is meant to produce thrust 
that will balance the drag on the hull of a ship, so $T_nU$ is a reasonable measure of useful power output.
Similarly, for a fish doing a C-start or escape manoeuvre \citep{domenici_kinematics_1997, liu_flow_2011}, its goal 
is to accelerate, such that $\eta_n$ is still a reasonable measure of efficiency that quantifies how much work is needed to attain a certain speed in a given amount of time. However, once the cruising speed $U$ is reached and the body moves at constant speed, $T_n=0$, and hence $\eta_n=0$. As pointed out by \citet{schultz_power_2002} among others, ``unless a fish is trying to `stir up the water,' it performs no useful work'' when swimming at constant speed. 

Indeed, at constant speed, the role of the propeller (for a ship) or of the swimming motion (for a fish) is to {\em compensate} the drag such as to keep the cruising velocity $U$. In an ideal fluid, there would be no drag on the body and no work would be needed to sustain velocity $U$: gliding would be enough. However, since water is a real fluid, if the fish was not swimming, or the propeller not rotating, the body would lose kinetic energy at a rate of:
\begin{equation}
P_{loss}=\frac{{\mathrm d}}{{\mathrm d} t}\left( \frac{1}{2}m\,U^2  \right) =-R U<0,
\end{equation} 
where $R$ is again the towed resistance at speed $U$ without a propeller (or a swimming motion). The goal of the propeller or of the swimming motion is to prevent this loss of kinetic energy due to the drag on the rigid body. Since the goal in this case is to compensate for the resistance $R$ and prevent the kinetic energy loss $P_{loss}$, a reasonable definition of useful power is:
\begin{equation}
P_{use}=P_{out}-P_{loss}=(T_n+R) U,
\end{equation} 
which we use to generalize the quasi-propulsive efficiency $\eta_{QP}$ to cases where the net thrust is not $0$:
\begin{equation}
\label{eq:eta_QP2}
\eta_{QP}=(T_n+R) U/P_{in}.
\end{equation} 
For the case of a self-propelled body moving at constant speed, $T_n=0$, and the definition of propulsive efficiency proposed in Eq. \ref{eq:eta_QP2} is the same as defined in Eq. \ref{eq:eta_QP}. The power $P_{in}$ is either experimentally
measured, or evaluated numerically as the time average of the power
to actuate the body. Finally, since towed experiments or simulations are often preferred to self-propelled ones for practical reasons, we will show in section \ref{} that Eq. \ref{eq:eta_QP2} can provide good estimates of the self-propelled quasi-propulsive efficiency under towed conditions.

In summary, three efficiencies (or quasi-efficiencies) can be defined to characterize the performance of a self-propelled body:
\begin{description}
\item[The net propulsive efficiency $\eta_n=T_nU/P_{in}$] \hfill \\
 uses the total net thrust of the \{body + propeller\} system. This efficiency is the one used to measure the performance of an isolated propeller. It is also a useful measure of the efficiency of acceleration manoeuvres. However, since $T_n=0$ in steady motion, it becomes meaningless when the goal of the system is not to accelerate or produce thrust.
 
\item[The propulsor efficiency $\eta_p=T_pU/P_{in}$] \hfill \\
uses the propulsor thrust. This is a measure of the hydrodynamic efficiency of the propeller that does not take into account the body-propulsor hydrodynamic interactions. Since there is no general way of separating the propulsor thrust from the body drag, especially for flexible bodies, estimating $T_p$ usually relies on more or less arbitrary models. For instance, in undulating swimming, $T_p$ is often estimated using an inviscid method such as Lighthill's elongated body theory \citep{lighthill_note_1960, lighthill_large-amplitude_1971}, or by separating the positive longitudinal forces from the negative ones \citep{borazjani_numerical_2008}.

\item[The quasi-propulsive efficiency $\eta_{QP}=(R+T_n)U/P_{in}$] \hfill \\
uses the resistance of the rigid body towed at speed $U$ without a propulsor. For a given vehicle and speed, the propulsion system with largest quasi-propulsive efficiency will minimize the energy consumption of the vehicle in steady motion. 
\end{description}


\section{Quasi-Propulsive Efficiency in Naval Architecture \label{sec:naval}}

In Naval Architecture, the use of the quasi-propulsive efficiency is
standard \citep{comstock_principles_1967} -- and straight-forward to use because the
body is rigid. Usually, to estimate the required power $P_{in}$, one
uses the propeller characteristics as measured in {\em open water},
i.e. with the propeller tested in isolation, without a hull in
front. The interaction between hull and propeller is accounted for
through factors derived either empirically or through additional
experimental tests.

The resistance of the ship under self-propelled conditions,
$R_{sp}$, will in general be larger than the towed resistance,
because the stern stagnation pressure (which is beneficial, reducing
the drag) is reduced due to the presence of the propeller which
accelerates the flow.  $R_{sp}$ is related to the towed resistance
$R$ through the``thrust deduction factor'' $t$, which depends on the
hull characteristics, the propeller characteristics and, primarily,
the hull-propeller interaction:
\begin{equation}
\label{eq3} R_{sp} = \frac {R}{1-t}.
\end{equation}
The factor $t$ is usually positive, reflecting the expectation that
the self-propelled resistance is larger than the towed resistance;
there may however be some cases where the reverse occurs, for hulls which
are bluff, i.e. not well streamlined, because of a reduction in the
separation effects. Another, but physically incorrect way to view
relation (\ref{eq3}) is that the propeller thrust $T$, which must be
equal to $R_{sp}$ in order to achieve self-propulsion, is {\em
reduced} when the propulsor is placed behind the vehicle, hence the
name thrust deduction.

Finally, since the propeller operates inside the wake of the
vehicle, the oncoming velocity is reduced compared to the free
stream velocity $U$; an averaged incoming velocity is used, $U_A$:
\begin{equation}
\label{eq4} U_A = U (1-w),
\end{equation}
where $w$, the ``wake fraction'', is derived empirically or with
separate experiments.  Hence, the useful power of the propeller must
be equal to $R_{sp} U_A$ in order to drive the vehicle in
self-propulsion.  If the propeller efficiency has been measured to
be equal to $\eta_p$ under ``open water conditions'', i.e. separately
from the vessel, then the input power must be equal to the useful
power divided by the propeller efficiency:
\begin{equation}
\label{eq5} Q_{in} = \frac{R_{sp} U_A}{\eta_p} = \frac{R U
(1-w)}{\eta_p (1-t)}.
\end{equation}
Substituting in equation (\ref{eq:eta_QP}) one finds:
\begin{equation}
\label{eq6} \eta_{QP} = \eta_p \frac{1-t}{1-w}.
\end{equation}
Finally, a factor is required to account for Reynolds number effects
on the propeller torque, caused by testing in model scale and in
uniform flow, the so-called ``relative propulsive efficiency''; but
this factor is not essential and we will not pursue it in the
present discussion.

As seen in equation (\ref{eq6}), the quasi-propulsive efficiency is
the product of the propeller efficiency  $\eta_p$ and the so-called hull
efficiency, $\eta_H$, defined as:
\begin{equation}
\label{eq7} \eta_{H} = \frac{1-t}{1-w},
\end{equation}
which accounts for the hydrodynamic interference between the hull
and the propeller.  Then, equation (\ref{eq6}) turns into:
\begin{equation}
\label{eq8} \eta_{QP} = \eta_p  \eta_H.
\end{equation}
It is usual, for example to have $\eta_{QP} > \eta_p$; and is even
possible, albeit rare, that $\eta_{QP}>1$ if the factors $t$ and $w$
reflect a large, favorable overall hydrodynamic interference.
Equation (\ref{eq8}), then, explicitly relates the system efficiency
($\eta_{QP}$) to the efficiency of the propulsor ($\eta_p$),
correcting for possible hydrodynamic interference ($\eta_H$).


\section{Quasi-Propulsive Efficiency in Fish Propulsion \label{sec:fish}}

For fish and fish-like propulsion, the problem is also rationally
defined using the {\em quasi-propulsive efficiency}, because what
should be important in terms of the energetics of a certain fish is
to employ a swimming mode that minimizes the power needed for
propulsion; whether this mode is hydrodynamically ``efficient'' is
secondary. Indeed, hydrodynamic efficiency suffers from the same
disadvantages mentioned in sections \ref{sec:intro}\&\ref{sec:eta_QP}, 
i.e. it is usually
impossible to separate drag from thrust, while a high propulsive
efficiency is meaningless, in terms of the stated objective to
minimize power expended, when the swimming mode causes adverse
hydrodynamic interference and actually increases the power needed.

The quasi-propulsive efficiency is again defined as in Eq. (\ref{eq:eta_QP}), where 
the power $P_{in}$ is either experimentally
measured, or evaluated numerically as the time average of the power
to actuate the body. The resistance $R$ is the {\bf drag on the fish
body towed at constant speed} $U$ and {\bf without bending}, i.e. the
resistance of a ``dead fish'' towed straight. Since similarly to the approach used in naval architecture, towed experiments or simulations are often preferred to self-propelled ones for practical reasons, we will show that Eq. (\ref{eq:eta_QP2}) can provide good estimates of the self-propelled quasi-propulsive efficiency under towed conditions.

\begin{figure}
 \centerline{\includegraphics[width=0.9\textwidth]{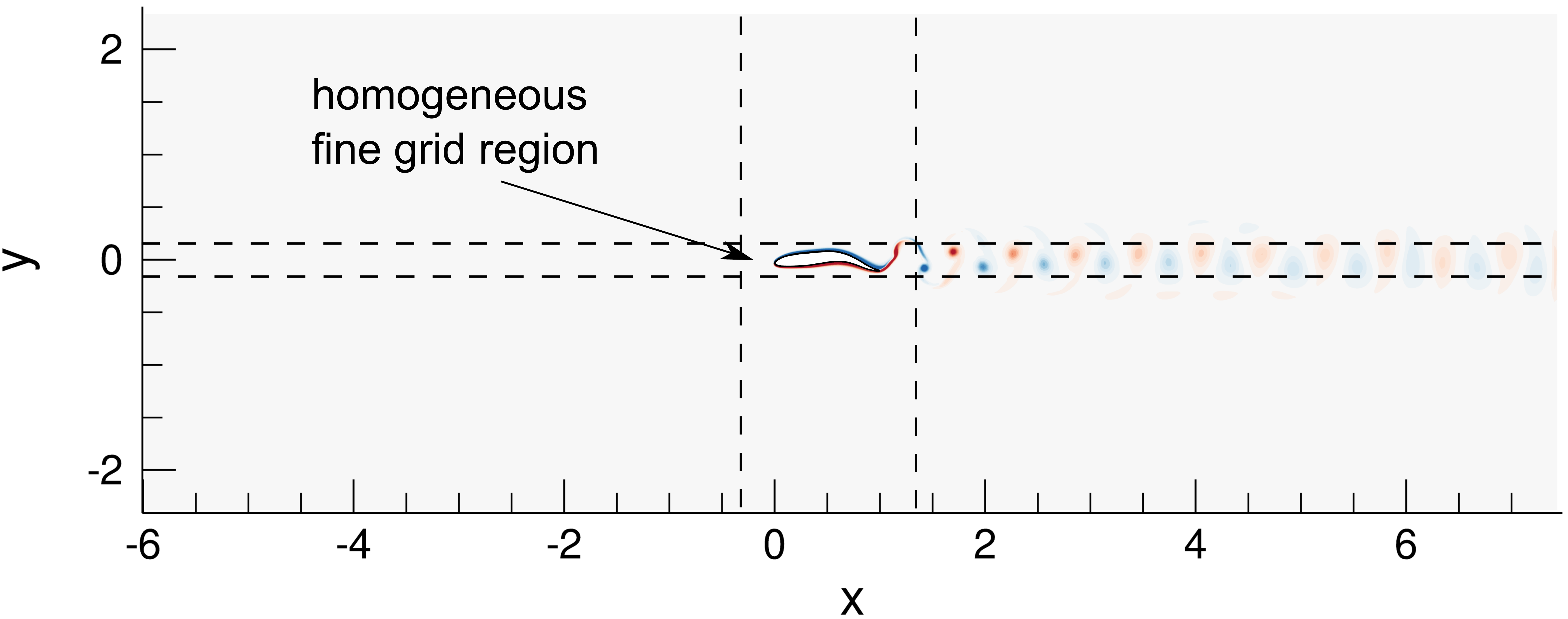}}
  \caption{Flow configuration for the BDIM simulations. The Cartesian grid is uniform near the undulating NACA0012 with grid size $\dd x=\dd y=1/160$ and uses a $1\%$ geometric expansion ratio for the spacing in the far-field. Constant velocity $u = U$ is used on the inlet, periodic boundary conditions on the  upper and lower boundaries, and a zero gradient exit condition with global flux correction. The vorticity field for the carangiform motion with $f=1.8$ and zero mean drag is shown as an example.}
\label{fig:geometry}
\end{figure}
In order to illustrate the discussion above, we will show through an example why
the quasi-propulsive efficiency is the only meaningful way of measuring propulsive
efficiency for self-propelled fishes or vehicles. In this example, the vehicle is represented by a self-propelled two-dimensional
 undulating NACA0012 foil of length $L$ swimming at average velocity $U$, chosen such that the Reynolds number is $Re=UL/\nu=5000$ (unless specified otherwise), where $\nu$ is the fluid dynamic viscosity. All lengths are normalized by $L$, velocities by $U$ and times (resp. frequencies) by $L/U$ (resp. $U/L$). The deformation of the foil is prescribed, while its heave and pitch motion are caused by the  hydrodynamic forces. These forces were estimated through two-dimensional viscous simulations on a Cartesian grid using the boundary data immersion method (BDIM) described \citet{weymouth_boundary_2011} and \citet{maertens_accurate_2014} on a domain represented in Figure \ref{fig:geometry}. 
 
The leading edge of the foil is located at $x=0$ and its trailing edge at $x=1$. The lateral displacement $h(x,t)$ of a point located at $x$ along the foil is  given at time $t$ by:
 \begin{align}
 \label{eq:motion}
 h(x,t) &= h_0(x,t)+B(x,t)  \nonumber \\
 &=\alpha A(x)\sin \big( 2\pi ( x/\lambda - ft )\big) + B(x,t)  \nonumber \\
 &=g(x)\cos \big( 2\pi (ft + \psi(x) )\big)
\end{align}  
where 
\begin{equation}
A(x) = 1+(x-1)c_1+(x^2-1)c_2 
\end{equation}
is the envelop of the prescribed travelling wave of wavelength $\lambda$ and frequency $f$, and 
\begin{equation}
B(x,t) = \left(a_r+b_rx \right) \sin \big( 2\pi (ft +\phi _r)\big)
\end{equation}
is the recoil term due to the hydrodynamic forces on the foil. $\alpha$ is the amplitude of $h_0$ at the trailing edge. It will either be kept constant ($\alpha =0.1$) or adjusted through a feedback control loop to ensure that the average drag on the foil is $0$.

\begin{figure}
	\centering
\begin{subfigure}[b]{0.47\textwidth}
	\centering
	\includegraphics[width=\textwidth]{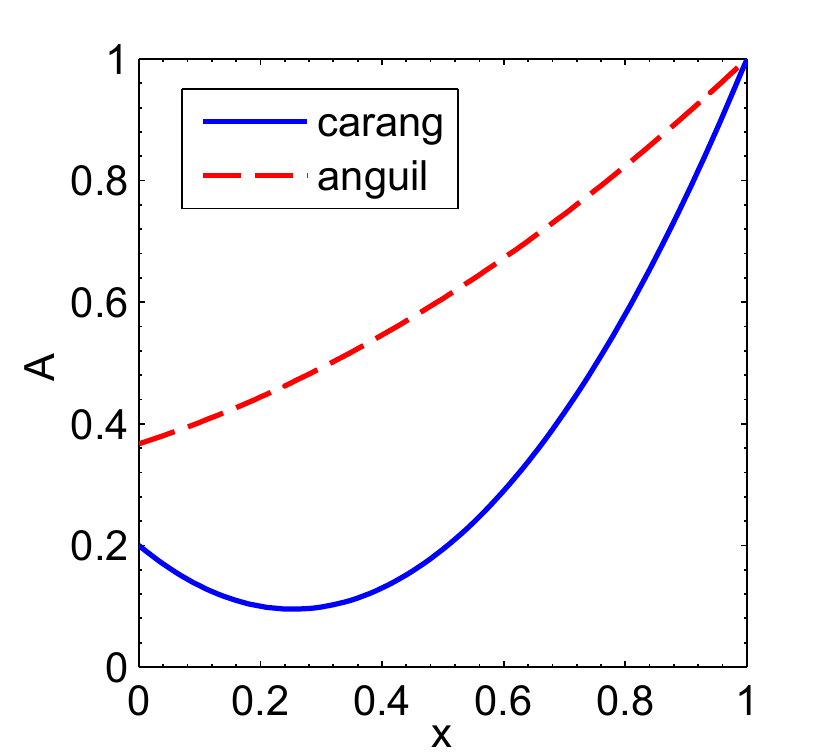}
	\caption{Prescribed amplitude envelopes}
	\label{fig:envelop}
\end{subfigure}
\hfill
\begin{subfigure}[b]{0.52\textwidth}
	\centering
	\includegraphics[width=\textwidth]{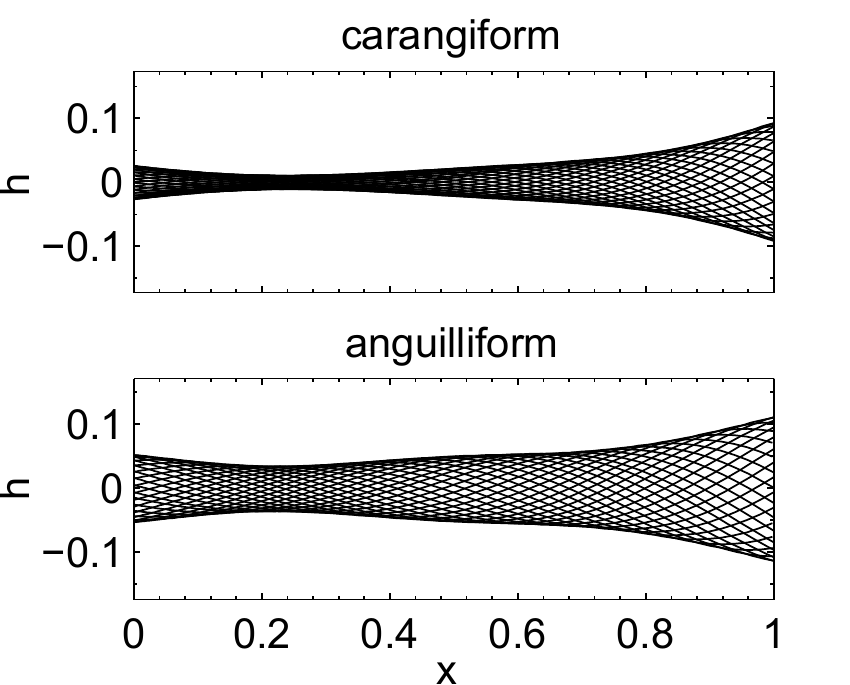}
	\caption{Midline displacement}
	\label{fig:motion}
\end{subfigure}
\caption{Carangiform and anguilliform motion for $f=1.8$ and $\alpha =0.1$.}
\end{figure}

Two envelops, represented in figure \ref{fig:envelop}, will be compared. The first one, widely used to represent carangiform gaits \citep{videler_fast_1984}, is characterized by:
\begin{equation}
\mathrm{carangiform:} \qquad c_1 = -0.825, \qquad c_2=1.625.
\end{equation}
The second one, representative of an anguilliform swimmer \citep{tytell_hydrodynamics_2004-1}, has parameters:
\begin{equation}
\mathrm{anguilliform:} \qquad c_1 = 0.323, \qquad c_2=0.310.
\end{equation}
The efficiency of the carangiform and anguilliform gaits in a towed and self-propelled configuration will be compared. $\lambda=1$ is used for all cases, while $f$ is varied in order to identify the \emph{most efficient} undulating frequency for both gaits.

\begin{figure}
 \centerline{\includegraphics[width=1\textwidth]{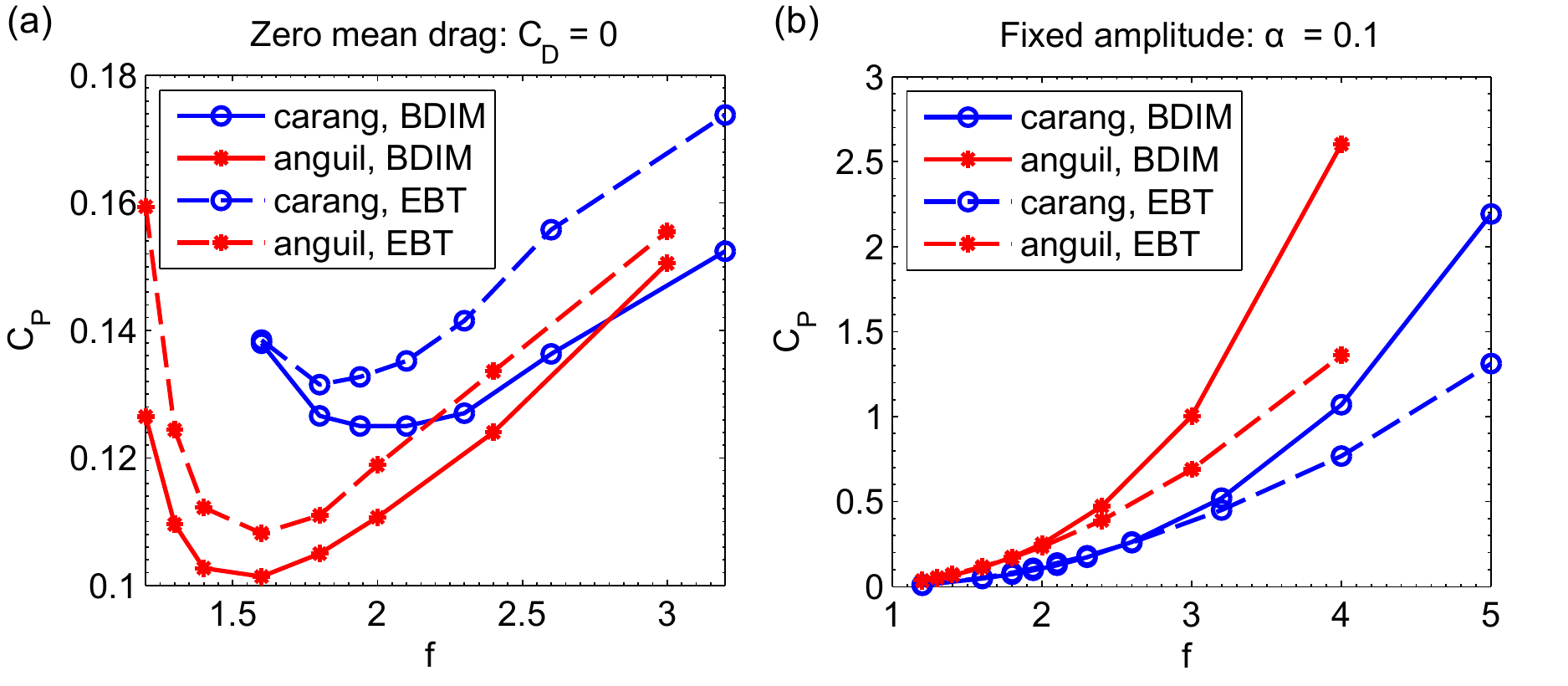}}
  \caption{Time-averaged power coefficient as a function of undulating frequency for (a) the zero drag and (b) the fixed amplitude configurations.}
\label{fig:CP}
\end{figure}

As mentioned above, the viscous simulation provides an estimate of the swimming power:
\begin{equation}
\label{eq:power}
P_{in}=\oint_{\partial B}\mathbf{v \cdot f^h} \,\mathrm{d}s
\end{equation}
and of the net thrust:
\begin{equation}
\label{eq:thrsut}
T_n=\oint_{\partial B} -f_x^h \,\mathrm{d}s
\end{equation}
where ${\mathbf f^h}$ are the hydrodynamic forces on the foil (with $x$-component $f^h_x$), ${\mathbf v}$ the local velocity of the undulating foil (as given by Eq. \ref{eq:motion}) and $\partial B$ the surface of the foil. From these values, we define the dimensionless power coefficient $C_P$ and thrust coefficient $C_T$:
\begin{equation}
C_P = \frac{P_{in}}{\frac{1}{2}\rho U^3L} \qquad \mathrm{and} \qquad  C_T = \frac{T_n}{\frac{1}{2}\rho U^2L}
\end{equation}
where $\rho$ is the fluid density. We similarly define the drag coefficient $C_D=-C_T$, as well as the friction ($C_{Df}$) and pressure ($C_{Dp}$) drag coefficients such that $C_D=C
_{Df}+C_{Dp}$.

Figure \ref{fig:CP}a shows that the self-propelled undulating NACA0012 foil travels with the least energy when using the anguilliform gait with frequency $f=1.6$, in which case $C_P=0.10$. If the carangiform gait was chosen, the most \emph{efficient} frequency would be $f=2$ with a power coefficient of $C_P=0.13$.
Though dimensionless, the power coefficient is not an intuitive measure of efficiency and does not allow easy comparison of various geometries.

Using the prescribed undulation and the recoil $B(x,t)$ calculated by the viscous simulation, Lighthill's Elongated Body Theory (EBT) also estimates the input power quite accurately \citep{lighthill_note_1960}, as can be seen on Figure \ref{fig:CP}. Moreover, unlike the viscous BDIM simulation, the EBT theory can provide an estimate of the propulsor thrust $T_{ebt} \approx T_p$ (see derivation in Appendix). Using the input power $P_{in}$ and the net thrust $T_n$ estimated from the BDIM simulation, as well as the EBT thrust and power estimates, we will now compare the efficiency of the various parameters estimated by the three measures defined in Section \ref{sec:efficiencies}.


\subsection{Net propulsive efficiency}
\begin{figure}
 \centerline{\includegraphics[width=0.6\textwidth]{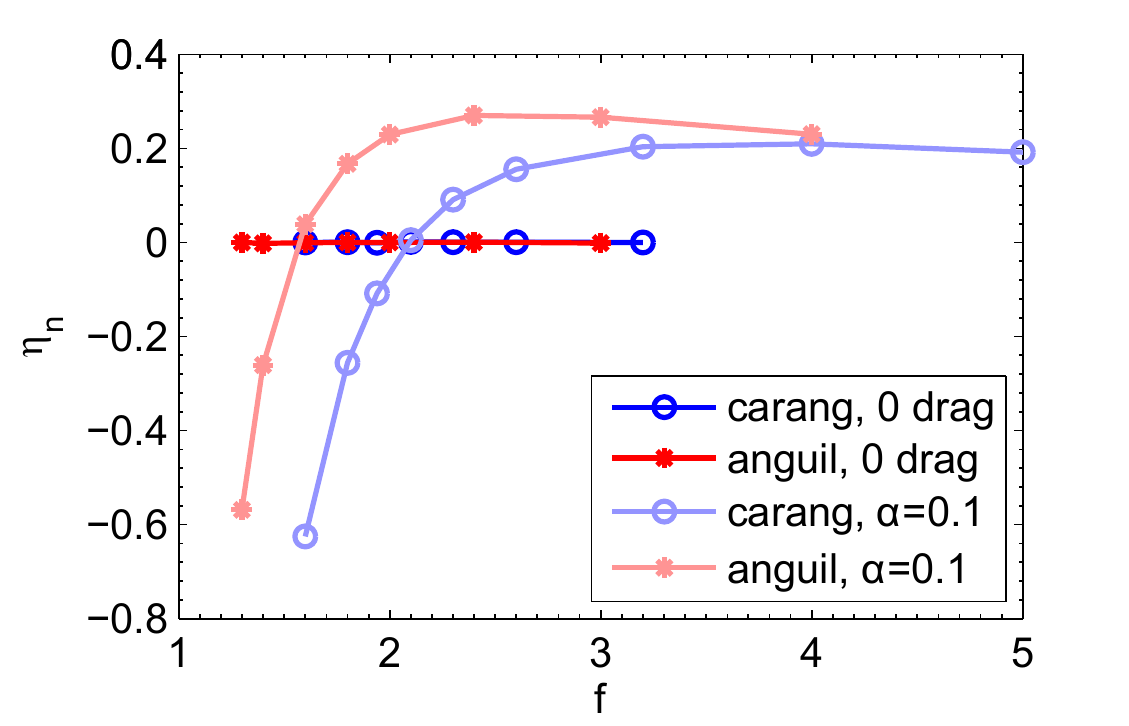}}
  \caption{Net propulsive efficiency}
\label{fig:eta_n}
\end{figure}

As discussed in previous sections, the net efficiency $\eta_n=T_nU/P_{in}$ is zero when the mean drag on the foil is $0$, which is the case for the self-propelled cases in Figure \ref{fig:eta_n}. It is therefore impossible to compare the performance of the two gaits or of the various frequencies using $\eta_n$ in these cases. As soon as the mean drag is non zero in the towed simulations ($\alpha =0.1$), it becomes clear that the anguilliform undulation is more efficient than the carangiform, but with values ranging from $-0.6$ to $0.3$, these undulating foils seem to be very poor propellers.

It is interesting to notice that, at low frequency, the net efficiency is negative due to a net drag on the undulating foil. What is the meaning of this negative efficiency? If we were considering a propeller, a net drag on the propeller would be counter productive and the ship might ``perform'' better without the propeller, so one intuitively expects the efficiency to be negative. However, in the case of a self-propelled undulating foil, an undulation that is not enough to overcome the drag, but it is able to reduce it, is not counter productive. Therefore, one would intuitively expect the efficiency to be positive. The quasi-propulsive efficiency solves this paradox by offering a measure of efficiency that is compatible with intuition.   

Now, if the goal is to accelerate the foil, a net thrust is needed. According to the net propulsive efficiency, the optimal undulating frequency is around $f=2.5$ ($\eta_n=0.27$) for the anguilliform motion and  $f=3.5$ ($\eta_n=0.21$) for the carangiform motion. These frequencies minimize the work required to attain a given acceleration. However, once the cruising speed has been reached and the goal is to minimize the power spent swimming in steady state, there is no guarantee that these frequencies are optimal. Indeed, these \emph{optimum} frequencies are different from those selected from figure \ref{fig:CP}.


\subsection{EBT propulsor efficiency \label{sec:ebt}}
\begin{figure}
 \centerline{\includegraphics[width=0.6\textwidth]{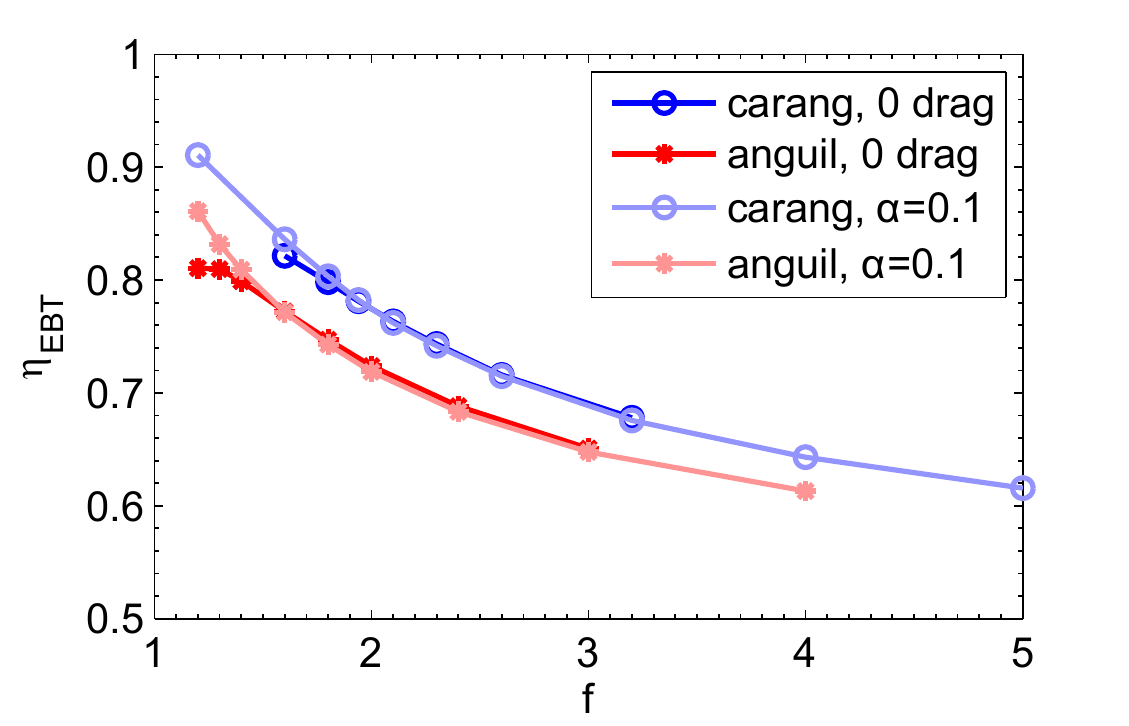}}
  \caption{Propulsor efficiency estimated from Lighthill's elongated body theory}
\label{fig:eta_EBT}
\end{figure}

In order to calculate the hydrodynamic efficiency of the undulating foil in the stationary regime, the thrust produced by the swimming motion needs to be estimated independently of the drag on the foil. This thrust can for example be estimated by one of the numerous inviscid methods. Here we use Lighthill's EBT \citep{lighthill_note_1960} which has a very simple expression (derived in Appendix) and proved very accurate at estimating the power. The dependency of $\eta_{ebt}=T_{ebt}U/P_{ebt}$ on the undulating frequency $f$, shown in Figure \ref{fig:eta_EBT}, leads to the conclusion, already drawn by Lighthill himself, that the lower the frequency the higher the hydrodynamic efficiency \citep{lighthill_large-amplitude_1971}.

\begin{figure}
 \centerline{\includegraphics[width=0.6\textwidth]{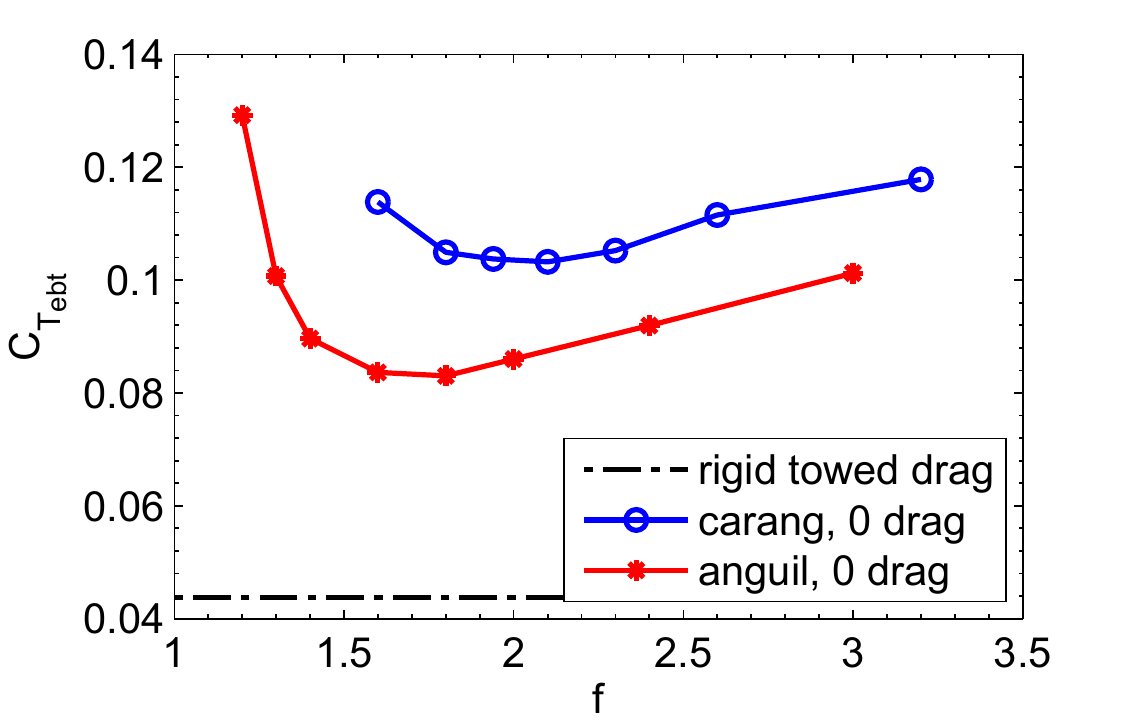}}
  \caption{Developing thrust estimated from EBT and comparison with the rigid towed drag.}
\label{fig:C_T}
\end{figure}

This result is, however, very deceptive. Indeed, the increase in efficiency at low frequency is not the result of a reduction in power consumption, but is due to an increase in the developing thrust, as can be seen in Figure \ref{fig:C_T}. For instance, there is a $15\%$ increase power consumption from $f=1.6$ to $f=1.3$, but in the same time, there is also a $20\%$ increase in thrust, resulting in a $5\%$ increase in propulsor efficiency. Since in both cases the net drag on the undulating foil is $0$, an increase in developing thrust must be balanced by an equal increase in drag on the body. So if $f=1.3$ seems hydrodynamically more efficient, $f=1.6$ is a better choice for the swimmer as it requires less work. Similarly, Figure \ref{fig:eta_EBT} suggests that the carangiform motion is more efficient than the anguilliform one, but this is purely due to a larger thrust, as Figure \ref{fig:CP} shows that the anguilliform motion is less energy intensive. 

This approach overestimates the efficiency by rewarding high thrust, which is also synonym of high drag. The ratio between the drag on the rigid towed body and the thrust produced by the swimming motion measures the hydrodynamic interference between the undulation and the body. The results presented in this section show that this factor, which is similar to the hull efficiency in naval architecture, plays a very important role: it is necessary to take it into account when comparing the efficiency of propulsion systems. 


\subsection{Quasi-propulsive efficiency}
\begin{figure}
 \centerline{\includegraphics[width=1\textwidth]{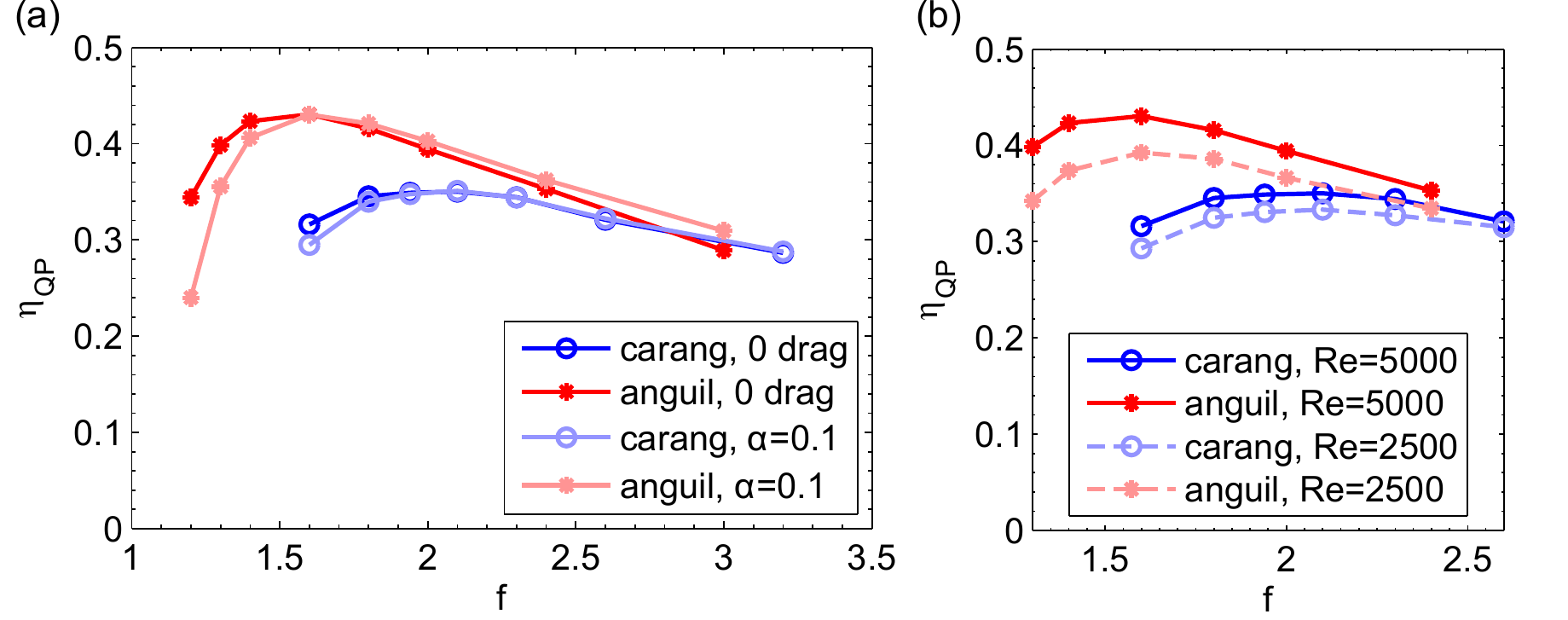}}
  \caption{Quasi-propulsive efficiency. (a): Comparison of towed estimates with self-propelled values ($Re=5000$). (b): Comparison of efficiency for $Re=2500$ and $Re=5000$ (self-propelled). }
\label{fig:eta_QP}
\end{figure}

Finally, $\eta_{QP}=(R+T_n)U/P_{in}$, with values comprised between $0.2$ and $0.5$, provides an intuitive and meaningful measure of the efficiency for the two undulating gaits at the various frequencies. Figure \ref{fig:eta_QP} shows that the carangiform gait, requiring less power, is an energetically better choice for a cruising undulating foil, and the best frequency is $f=1.6$ with an efficiency of $43\%$. For the carangiform undulation, the maximum efficiency drops to $35\%$ for the frequency $f=2.1$.

Since self-propelled experiments and simulations are often more challenging than towed ones, it is of high practical interest to be able to estimate the quasi-propulsive efficiency from towed experiments. Figure \ref{fig:eta_QP}a also shows that the estimates obtained by keeping the amplitude $\alpha$ constant instead of ensuring $0$ mean drag are very close to the self-propelled values (except at the very low frequencies). 

Within the same hydrodynamic regime, the values of $\eta_{QP}$ for different Reynolds numbers are also of comparable amplitude, on a natural unit scale. For instance, figure \ref{fig:eta_QP}b compares the efficiency of the same self-propelled undulating motion for two different Reynolds numbers: $Re=2500$ and $Re=5000$. Even though the power coefficient increases by $50\%$ from $Re=5000$ to $Re=2500$, the difference in efficiency between the two Reynolds numbers is no more than $7\%$ and their trends are very similar. This result therefore corroborates what the intuition would expect: within a given hydrodynamic regime, the efficiency only weakly depends on the Reynolds number. This 
also illustrates that, even though both $C_P$ and $1/\eta_{QP}$ are normalized versions of the swimming power, $C_P$ is not very convenient to use due to its strong dependence on Reynolds number.

Finally, we would like to remark that, as the thrust produced by the undulating foil increases, $\eta_{QP}$ converges to $\eta_n$. Indeed, if $T_n \gg R$, then $\eta_{QP} \approx T_nU/P_{in}$. Since this is typically the case for a propeller, the drag on the hull being much larger than that of the propeller, $\eta_{QP}$ can be seen as a generalization of the traditional propeller efficiency to the low thrust regime.


\subsection{Remarks on the efficiency of undulating swimming}

With the undulation gaits used in this example, the propulsor efficiencies are comparable to typical ship propeller efficiencies, but the quasi-propulsive efficiencies attained are much smaller, by a factor of $2$ to $3$. As mentioned in Section \ref{sec:ebt}, this difference is explained by the fact that the thrust predicted by the EBT theory is at least twice as large as the towed drag, as shown on Figure \ref{fig:C_T}. Assuming the EBT estimate has the same error for thrust prediction as for power, namely less than $25\%$, this $100-200\%$ difference suggests that the hull efficiency is quite low for this propulsion system. In other words, body undulations, while producing thrust, are also responsible for a significant increase in drag.
 
This observation, which has often been made in the literature  , is at the core of a century long controversy opposing the drag reduction proponents \citep{gero_hydrodynamic_1952, fish_dolphin_1991, fish_passive_2006} in the wake of Gray and his famous \emph{paradox}  \citep{gray_studies_1936}, to the drag enhancement advocates \citep{lighthill_large-amplitude_1971, webb_hydrodynamics_1975, alexander_swimming_1977, videler_swimming_1981}. While the latter have long conjectured that body undulations must significantly increase the skin friction along the body due to what is often referred to as the `Bone-Lighthill boundary-layer thinning hypothesis' \citep{lighthill_large-amplitude_1971}, such an increase has never been confirmed. 
Instead, experimental visualization of the boundary layer of dead towed and live self-propelled fishes showed that the skin friction on a fish, undulating or not, was just higher than the drag on a flat plate (Anderson 2005). Similarly, theoretical analysis from \citet{ehrenstein_skin_2013} also suggested an increase in the skin friction drag on the order of $20\%$, well bellow the Bone-Lighthill hypothesis values of $3$ to $5$ \citep{lighthill_large-amplitude_1971}. 

\begin{figure}
 \centerline{\includegraphics[width=1\textwidth]{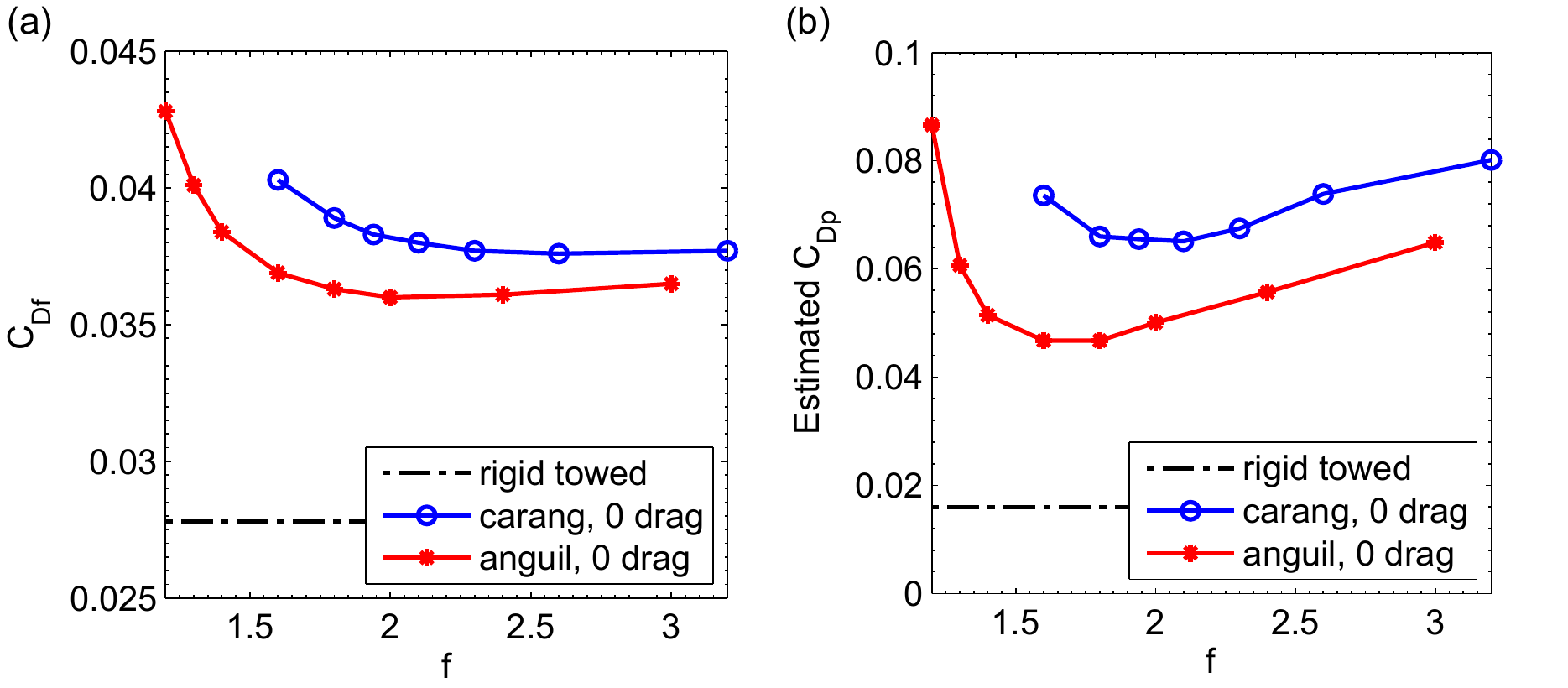}}
  \caption{(a) Friction drag on undulating foil compared to rigid towed friction drag. (b) Pressure drag estimated from BDIM friction drag and EBT thrust: $C_{Dp}=C_{T_{ebt}}-C_{Df}$.}
\label{fig:C_Df-p}
\end{figure}

Our simulations of undulating foils in which power, friction and pressure along the self-propelled foil are simultaneously estimated can help shed a new light on this controversy. Despite low swimming efficiencies and high EBT thrust, we found that when the body undulates, the friction drag increases by no more than $60\%$, as shown in Figure \ref{fig:C_Df-p}a. A necessary corollary is that the drag enhancement is primarily due to an increase in form drag, as shown in Figure \ref{fig:C_Df-p}b, by a factor of $3$ to $5$. This counter-intuitive result reinforces the notion that drag cannot be separated from thrust on an undulating body at $Re \gg 1$, as pressure is the primary source of both.

In the examples used in this paper, we found hull efficiencies around $30-50\%$ due to adverse hydrodynamic interactions between the foil and the undulating motion, resulting in a significant increase in form drag. However, experiments on a robotic tuna by \citet{barrett_drag_1999} suggest that, especially at high Reynolds number, it is possible
for the undulating motion to interact beneficially with the drag on the body and obtain quasi-propulsive efficiencies larger than $1$. \citet{barrett_drag_1999} measured directly the power needed to drive the
tuna-like motion of a robotic mechanism under self-propulsion
conditions.  Inviscid theory provided values for the self-propulsion
power very close to the experimentally measured values \citep{barrett_drag_1999, kagemoto_force_2000, smith_simulation_2004}. The
quasi-propulsive efficiency, estimated as proposed herein, provided
values up to $150\%$, well in excess of $100\%$, which simply means that
the resistance of the actively swimming body was less than the drag
under straight-towing conditions. The measurements were at the
transitional Reynolds number of around $Re=800\,000$ where
re-laminarization of the boundary layer and separation suppression
is possible. Indeed, simulations \citep{shen_turbulent_2003} and experiments
\citep{techet_separation_2003} on an actively flapping two-dimensional sheet
demonstrated clear turbulence reduction, in addition to flow
separation suppression, which was noted earlier by \citet{taneda_visual_1977}.
This explains the drop in drag under self-propulsion conditions and
hence the high quasi-propulsive efficiency values; indeed \citet{barrett_drag_1999} find the equivalent drag coefficient of the actively
swimming mechanism to be closer to laminar boundary layer values,
whereas the drag coefficient of the straight-towed mechanism is
close to turbulent boundary layer values.


\section{Discussion}

\subsection{Notion of drag/thrust on a self-propelled body}

\citet{schultz_power_2002} discussed the difficulty of establishing a
system propulsive efficiency for self-propelled bodies. They applied
the concept of propulsor efficiency to define the system efficiency;
since the net force is zero (as it must be in every self-propelled
body), the system efficiency defined in this manner is zero as well;
this is not a helpful result, because any system, however wasteful
its propulsor may be, will be deemed equally (in)efficient as any
other. 

The difficulty of establishing a propulsive efficiency stems from the impossibility to separate drag and thrust since they balance on average and pressure (resp. viscosity) is the primary source of both at large (resp. low) Reynolds number. Inviscid approaches propose thrust estimates, but these remain controversial due to the blurry definition of thrust for a self-propelled body. For instance, it is sometimes argued that Lighthill's model overestimates the thrust \citep{hess_fast_1984, anderson_boundary_2001, shirgaonkar_new_2009}. The quasi-propulsive efficiency moves away from the ill-defined notion of drag on a self-propelled body, using the well defined drag on a towed body instead. It results in an intuitive measure of efficiency that can be used to minimize the ``fuel'' consumption  rather than the hydrodynamic efficiency. 

Although the notion of thrust is ill-defined, attributing high (resp. low) quasi-propulsive efficiencies to a drag reduction (resp. enhancement) is a possible way of interpreting the performance of a propulsion system. If this interpretation is applied to the examples used in this paper, it is found that increases in the well defined friction drag
could not alone account for the low swimming efficiency, and a significant increase in pressure drag has to be hypothesised. It might, however, be possible for drag reduction mechanisms to compensate for these drag increase and result in highly (quasi-propulsive) efficient swimming gaits. Such mechanisms used by fish and mammals, either passive or active, are reviewed in \citet{fish_passive_2006}. 

\subsection{A universal measure of efficiency for swimmers}

Unlike propulsor efficiencies relying on inviscid thrust models, the quasi-propulsive efficiency is ass appropriate for low-Reynolds-number swimming motions as for large-Reynolds-number ones. 
\citet{becker_self-propulsion_2003} define and use a system efficiency which is the
same as the quasi-propulsive efficiency definition herein; they
study a three-link micro-propulsor, employing flexing of the links
to achieve locomotion at very low Reynolds numbers. In the words of
the authors, ``We define a swimming efficiency as the power necessary
to pull the straightened swimmer along its axis at the average speed
of the actual swimmer, relative to the average mechanical power
generated by the actual swimmer to achieve that speed.''  It is
important to note that the useful power is defined in terms of the
{\em towed straightened swimmer}.  In fact, for very low Reynolds
number it is impossible to distinguish thrust from drag, since
viscous forces produce both forces, making the use of the
quasi-propulsive efficiency essential.  Micro-swimmers have,
typically, less than a few percent efficiency.

\subsection{Optimizing velocity and  body shape \label{sec:general}}
We have shown through examples that quasi-propulsive efficiency $\eta_{QP}$ is the only rational measure of the efficiency for a self-propelled body in steady motion. There is no theoretical guarantee that $\eta_{QP}$ will be smaller than $1$, and it can indeed be greater than $1$ for very efficient propulsion \citep{barrett_drag_1999}. However, it gives an intuitively meaningful number that allows the comparison of various geometries and propulsion systems. It can, for instance, be used to compare the efficiency of man-made systems and biological ones. It can ot, however, be used to compare or optimize the performance of hull or body shapes \citep{kagemoto_why_2013, van_rees_optimal_2013}, or swimming velocities \citep{liu_optimal_2012}. 

A more general goal than that of Section \ref{sec:eta_QP} can be expressed as: {\em For a
given mass $m$, find the body shape, propulsor and velocity that will require the least amount of energy to drive the vehicle from point $A$ to point $B$ in a fluid of kinematic viscosity $\nu$ and density $\rho$.}

In other words, the goal is to minimize the energy per unit length travelled for a mass $m$ in a given fluid. For this problem, the natural units are:
\begin{equation}
\mathrm{mass:}\ m,\qquad
\mathrm{length:}\ \left(\frac{m}{\rho}\right)^{1/3},\qquad
\mathrm{time:}\ \nu \left(\frac{m}{\rho}\right)^{2/3}.
\end{equation}
If the average swimming power is $P_{in}$ and the average velocity is $U$, the average energy $E$ spent per unit length (using the length unit defined above) is:
\begin{equation}
E = \frac{P_{in} m^{1/3}}{U \rho ^{1/3}}.
\end{equation}
The corresponding dimensionless coefficient, which we will call energy coefficient $C_E$, is:
\begin{equation}
C_E = \frac{P_{in}}{\rho U \nu ^2}.
\end{equation}
Unlike the quasi-propulsive efficiency, this energy coefficient is convenient for comparing various geometries and propulsion strategies. However, $C_E$ is decreasing with Reynolds number, therefore any optimization would conclude that a swimming speed of zero is optimal since it does not require energy. Indeed, the coefficient $C_E$ takes into account the hydrodynamic power spent to travel from $A$ to $B$, but nothing ensures that the travel will be accomplished in a finite time. This difficulty can easily be overcome by adding a cost to time. From a biological point of view, $P_{in}$ can be replaced by the total power $P_t$ defined in \citet{liu_optimal_2012} as:
\begin{equation}\label{eq:P_t}
P_t=\frac{P_{w}}{\beta}+P_m,
\end{equation}
where $\beta$ is the muscle power efficiency, $P_m$ is
the standard metabolic rate independent of swimming speed and $P_{w}$ is the hydrodynamic power (similar to the definition of $P_{in}$ in Eq. \ref{eq:power}).

We now define the performance index:
\begin{equation}
C_{\eta}=\frac{\rho U \nu ^2}{P_t},
\end{equation}
that can be used to solve the very general problem of optimizing the body shape, swimming speed and propulsion system. It is very similar, in spirit, to Liu's energy-utilization ratio $f_{\eta}$ \citep{liu_optimal_2012}, but Liu normalized $f_{\eta}$ with parameters depending on the geometry ($L$) and propulsion ($\lambda$). Even though the performance index could also be used to solve the optimization problem presented in Section \ref{sec:eta_QP}, its order of magnitude varies widely with Reynolds number: $C_{\eta} \sim 10^{6}$ for the examples considered in this paper. The quasi-propulsive efficiency, with a natural scale going from $0$ to $1$, is much more intuitive and easy to work with. 

\subsection{Word of caution}
Finally, a word of caution: As \citet{fish_porpoise_2005} demonstrates in his
outline of the controversial {\em Gray's paradox}, estimates of the
propulsive parameters must be made very carefully, especially
because the quasi-propulsive efficiency is based on two separate
experiments, one for a towed body (numerator, or useful work) and
one for the self-propelled experiment (denominator, or expended
energy); both experiments must refer to the same speed and the same
duration of time over which performance is assessed, otherwise
erroneous conclusions may be drawn.

One should also keep in mind that the mechanical efficiency, considered in this paper, is only the last link in a series of processes involved in swimming. As  \citet{ellerby_how_2010} explains in his short review of Webb's contributions, `For fish, just as with engineered vehicles, fuel consumption is the most obvious measure of power input.' Fuel comes in the form of metabolic energy, and the efficiency of converting this chemical energy to mechanical energy plays an important role in the final measure of swimming efficiency, as hinted by the total power defined in equation \ref{eq:P_t}.


\section{Conclusion}

The optimal propulsor for a self-propelled system is the one that
minimizes fuel consumption for a given body size and speed.  The
hydrodynamic efficiency is not a good measure of optimality, because
the numerator (the useful energy) is not easily defined in fish,
since drag is difficult to measure, and, far more importantly, its
value depends on the propulsion mode employed. 

A non-dimensional quantity that can be used to minimize power
consumption and allows comparison of various systems at different
scales, is the quasi-propulsive efficiency, defined as the ratio of
the energy needed to tow the fish straight at a given speed divided
by the power to self-propel itself at the same speed. This is a
rational non-dimensional metric of the system propulsive fitness.


\section*{Appendix: Elongated Body Theory}
We call $h(x,t)$ being the lateral displacement of any cross-section of an undulating fish, the relative fluid velocity at any cross-section is given by:
\begin{equation} \label{eq:h}
w = \frac{\partial h}{\partial t} + U\frac{\partial h}{\partial x} .
\end{equation} 
If only the hydrodynamic force associated with the vortex sheet from the tail is considered then, according to the EBT \citep{lighthill_note_1960, lighthill_mathematical_1975, cheng_note_1994}, the time-averaged values of power required ($P_{ebt}$), thrust ($T_{ebt}$) and efficiency ($\eta_{ebt}$) are:
\begin{subequations} \label{eq:definitions}
\begin{empheq}[left=\empheqlbrace]{align}
&  P_{ebt} = \rho U A\, \overline{ \left( w\, \frac{\partial h}{\partial t} \right)} \\
&  T_{ebt} = P_{ebt}/U -\frac{1}{2}\rho \overline{w^2} A \\
&  \eta_{ebt} = \frac{T_{ebt}U}{P_{ebt}}    
\end{empheq}
\end{subequations}
In the small amplitude version of the theory, $U$ velocity in the longitudinal direction is assumed to be constant. From our experience, this is a fair approximation. $A$ is the added mass of the section per unit length $L$ (in 2D, $A=\rho \pi/4$). All the values are calculated at the trailing edge ($x=L$).

$h(x,t)$, can be expressed as:
\begin{equation}
h(x,t) = g(x)\cos \big( 2\pi (ft + \psi(x) )\big) = \Real (H(x,t)),
\end{equation}
where for convenience, we define the complex function:
\begin{equation}
H(x,t) = G(x)e^{\ic \omega t}.
\end{equation}
We then have:
\begin{equation}
\frac{\partial H}{\partial t} = \ic \omega G(x) e^{\ic \omega t}, 
\qquad \frac{\partial H}{\partial x} = G'(x) e^{\ic \omega t}. 
\end{equation}
We also define $G_0$ (resp. $G'_0$) the amplitude of the complex number $G(L)$ (resp. $G'(L)$), such that:  
\begin{equation}
G(L) = G_0 e^{\ic \theta}, \qquad G'(L) = G'_0 e^{\ic \alpha}.
\end{equation}
Using these complex functions:
\begin{equation} \label{eq:mean}
\overline{\left(\frac{\partial h}{\partial t} \right)^2} = \frac{1}{2}\omega^2 {G_0}^2,
\qquad \overline{\left(\frac{\partial h}{\partial x} \right)^2} = \frac{1}{2} {G'_0}^2,
\qquad \overline{\frac{\partial h}{\partial t} \frac{\partial h}{\partial x}} =\frac{1}{2} \omega G_0 G'_0 \sin (\alpha -\theta ).
\end{equation}
Substituting Eqs. \ref{eq:h} and \ref{eq:mean} into Eq. \ref{eq:definitions}, we have:
\begin{subequations}
\begin{empheq}[left=\empheqlbrace]{align}
& P_{ebt} = \rho U A \, \overline{\frac{\partial h}{\partial t} \left(\frac{\partial h}{\partial t} +U\frac{\partial h}{\partial x}   \right)} = \frac{1}{2} \rho U A \left(\omega ^2 {G_0}^2 + U \omega G_0 G'_0 \sin (\alpha -\theta )\right)  \\
&T_{ebt}U = \frac{1}{2}\rho U A \,\overline{\left[\left(\frac{\partial h}{\partial t}\right)^2 -U^2 \left(\frac{\partial h}{\partial x}\right)^2 \right]} = \frac{1}{4}\rho U A \left(\omega ^2{G_0}^2 -U^2 {G'_0}^2 \right) \\
& \eta_{ebt} = \frac{T_{ebt}U}{P_{ebt}}  = \frac{1}{2}\frac{\omega ^2{G_0}^2 -U^2 {G'_0}^2}{\omega ^2 + U \omega G_0 G'_0 \sin (\alpha -\theta )}
\end{empheq}
\end{subequations}
We now define $a=2G_0$ the double amplitude of the tail motion and the two variables:
\begin{equation}
 X = \frac{\omega G_0}{UG'_0} \qquad \text{and} \qquad \Delta = \alpha - \theta .
\end{equation}
The expression of the power and thrust simplify to:
\begin{subequations}
\begin{empheq}[left=\empheqlbrace]{align}
& P_{ebt} = \frac{1}{8} \rho U A \omega ^2 a^2 \left(1+\frac{1}{X} \sin (\alpha -\theta )\right)  \\
& T_{ebt}U =  \frac{1}{16}\rho U A \omega ^2 a^2 \left(1-\frac{1}{X^2} \right) 
\end{empheq}
\end{subequations}
The EBT efficiency only depends on $X$ and $\Delta$ and has the simple expression:
\begin{equation}
\eta_{ebt} = \frac{T_{ebt}U}{P_{ebt}}  = \frac{1}{2}\frac{X-1/X}{X+\sin (\Delta)}. 
\end{equation}

\bibliographystyle{abbrvnat}
\bibliography{fish}

\end{document}